\documentstyle[preprint,aps]{revtex}

\def\footnoterule{\kern-3pt \hrule width \hsize \kern6.2pt}

\def\pmb#1{\setbox0=\hbox{$#1$}%
\kern-.025em\copy0\kern-\wd0%
\kern.05em\copy0\kern-\wd0%
\kern-.025em\raise.0433em\box0}%

\def\be{\begin{equation}}
\def\ee{\end{equation}}
\def\beq{\begin{eqnarray}}
\def\eeq{\end{eqnarray}}

\def\pa{\partial}

\begin{document}

\renewcommand{\thefootnote}{\fnsymbol{footnote}}
\thispagestyle{empty}
\setcounter{page}{0}

\title{DERIVING THE HARD THERMAL LOOPS OF QCD\\
FROM CLASSICAL TRANSPORT THEORY\footnotemark[1]}

\footnotetext[1]{\baselineskip=12pt This
work is supported in part by funds
provided by the U.S.~Department of Energy (D.O.E.) under contract
\#DE-AC02-76ER03069. P.F.K. is supported by the Natural Sciences
and Engeneering Research Council of Canada. C.L. is supported by the
Swiss National Science Foundation. C.M. is supported by the
Ministerio de Educaci\'on y Ciencia, Spain.\hfill\break\break
Internet addresses:
kelly@pierre.mit.edu (P.F.K.),
liu@pierre.mit.edu (Q.L.),\hfill\break
$\phantom{Internet addresses: }$lucchesi@pierre.mit.edu (C.L.),
cristina@pierre.mit.edu (C.M.).\hfill\break}

\author{P.F.~Kelly,~~Q.~Liu,~~C.~Lucchesi and~~C.~Manuel}

\address{Center for Theoretical Physics,
Laboratory for Nuclear Science, and Department of Physics\\
Massachusetts Institute of Technology, Cambridge,
Massachusetts 02139}

\maketitle

\renewcommand{\thefootnote}{\arabic{footnote}}
\setcounter{footnote}{0}
\setcounter{page}{0}
\thispagestyle{empty}

\begin{abstract}
$\!\!$Classical transport theory is employed to analyze the hot
quark-gluon
plasma at the leading order in the coupling constant.
A condition on the (covariantly conserved) color current is obtained.
{}From this condition, the
generating functional of hard thermal loops
with an arbitrary number of soft external bosonic legs can be derived.
Our approach, besides being more direct
than alternative ones, shows that hard thermal loops are
essentially classical.
\end{abstract}

\vfill
%\centerline{Submitted to: {\em Phys. Rev. Lett.}}
%\vfill

\noindent
PACS No: 12.38.Mh, 51.10.+y, 11.10.Wx, 11.15.Kc
\hfill\break
\hbox to \hsize{CTP\#2292 \hfil March 1994}
\vskip-12pt
\eject

It is the purpose of the present Letter to derive
the hard thermal loops (HTL's) of QCD from classical transport theory.
The generating functional of HTL's (with an arbitrary number of soft
external bosonic legs) arises as a leading-order effect in this
formalism.

In the diagrammatic approach to high-temperature QCD it
was realized that a resummation procedure is necessary in order to
take into account consistently all contributions at leading-order in
the
coupling constant \cite{BPFT}. Such contributions arise from one-loop
diagrams with external soft legs and hard internal momenta,
the so-called ``hard thermal loops". The HTL approach provides
gauge invariant results for physical quantities.
An effective action for HTL's was given in \cite{TW}
by exploring the gauge invariance condition for the generating
functional.
This condition has been identified with the equations of
motion for the topological Chern-Simons theory at
zero temperature \cite{EN}. Furthermore, the eikonal for a
Chern-Simons theory has been used to obtain a non-Abelian
generalization of the
Kubo formula \cite{JN}, which governs, through the current induced
by HTL's, the response of a hot quark-gluon plasma.

Another successful description of the hard thermal loops in QCD
is based on
(a truncation of) the hierarchy of Schwinger-Dyson equations
\cite{BI}; the generating functional for HTL's is obtained by
performing a consistent expansion in the coupling constant.
Alternatively, one can start from the composite effective action
\cite{CJT} by requiring its stationarity \cite{JLL}, and by using the
approximation scheme developed in \cite{BI}.
The gauge invariance condition for the
generating functional can then be obtained as a consequence of symmetry
properties due to the form of the momentum integration measure (see
\cite{JLL}).

The diagrammatic techniques of \cite{BPFT}, as well as the
consistent expansion in the coupling constant presented in \cite{BI},
although remarkably insightful,
are technically very involved and necessitate lengthy
computations. Furthermore, they are puzzling in respect of the very
nature of hard thermal loops. One wonders to what extent these are
quantum effects: on the one hand HTL's emerge in loop
diagrams and are described by the Schwinger-Dyson equations
of quantum field theory. On the other hand it is generally believed
that hard thermal effects are UV-finite since they are due
exclusively to {\it thermal} scattering processes.

This motivates us to find an alternative, {\it classical} formalism for
hard thermal loops in QCD. The obvious setting for such a search is the
classical transport theory of plasmas (see for instance \cite {LL}). Our
effort was encouraged by the fact that for an Abelian plasma of
electrons and ions, the dielectric tensor that can be computed
\cite{Silin}
from classical transport theory is the same as the one that can be
extracted from the hard thermal corrections to the
vacuum polarization tensor \cite{BPFT,JN}. Moreover, the same situation is
encountered for non-Abelian plasmas \cite{EH,Weldon}.
The transport
theory for colored plasmas has been established in \cite {EH}. There
has not been, to the best of our knowledge, any
attempt at deriving the complete set of HTL's from this formalism.
It is our purpose to do so, for the HTL's with an arbitrary number of
external soft bosonic legs.

Consider a particle bearing a non-Abelian $SU(N)$ color charge
$Q^{a}, \ a=1,...,N^2-1$, traversing a worldline $x^{\alpha}(\tau)$,
where $\tau$ denotes the proper time.
The dynamical effects of the spin of the particles shall be ignored, as
they are typically small. The Wong equations
\cite{Wong} describe the dynamical evolution of $x^{\mu}$,
$p^{\mu}$ and $Q^{a}$:
\begin{mathletters}
\label{wongeq}
\beq
m\, {{d x^{\mu}}\over{d \tau}} & = & p^{\mu}
\ , \label{wongeqa} \\[2mm]
m\, {{d p^{\mu}}\over{d \tau}} & = & g\,Q^{a}F^{\mu\nu}_{a}p_{\nu}
\ ,\label{wongeqb}\\[2mm]
m\, {{d Q^{a}}\over{d \tau}} & = & - g\, f^{abc}p^{\mu}A^{b}_{\mu}Q^{c}
\ . \label{wongeqc}
\eeq
\end{mathletters}
$\!\!$The $f^{abc}$ are the structure constants of the group,
$F^{\mu\nu}_{a}$ denotes the field strength, $g$ is the coupling constant,
and we set $c=\hbar=k_B=1$ henceforth.
Equation (\ref{wongeqb}) is the non-Abelian generalization of the
Lorentz force law, and eq. (\ref{wongeqc}) describes the precession
in color space of the charge in an external color field $A_\mu^a$. It is
noteworthy that the color charge is itself subject to
dynamical evolution, a feature which distinguishes the non-Abelian theory
from electromagnetism.

The  usual $(x,p)$ phase space is enlarged by including color degrees of
freedom for colored particles.
Physical constraints are enforced by  the choice
of the volume element $dx\,dP\,dQ$. The momentum measure
\begin{equation}
dP = {{d^{4}p}\over{(2\pi)^{3}}}\,\,2\,\theta(p_{0})\,\,\delta(p^{2} - m^{2})
\label{measurep}
\end{equation}
guarantees on-shell evolution and positivity of the energy.
The color charge measure enforces the conservation
of the group invariants, {\it e.g.}, for $SU(3)$,
\begin{equation}
dQ = d^8 Q\,\, \delta(Q_{a}Q^{a} - q_{2})\,\,
\delta(d_{abc}Q^{a}Q^{b}Q^{c} - q_{3}) \ , \label{measureQ}
\end{equation}
where the constants $q_{2}$ and $q_{3}$ fix the values of the Casimirs.

The one-particle distribution function $f(x,p,Q)$ is the classical
probability density for finding the particle in the state $\{x,p,Q\}$.
This probability density evolves in time via a transport equation,
\begin{equation}
m\, {{d f(x,p,Q)}\over{d \tau}} = C[f](x,p,Q) \ , \label{transport}
\end{equation}
where $C[f](x,p,Q)$ denotes the collision integral, which we henceforth
set to zero. Using the equations of
motion~(\ref{wongeq}), in the collisionless
case, (\ref{transport}) becomes the Boltzmann equation,
\begin{equation}
p^{\mu}\left[{{\partial}\over{\partial x^{\mu}}}
- g\, Q_{a}F^{a}_{\mu\nu}{{\partial}\over{\partial p_{\nu}}}
- g\, f_{abc}A^{b}_{\mu} Q^{c}{{\partial}\over{\partial Q_{a}}}
\right] f(x,p,Q) = 0 \ . \label{boltzmann}
\end{equation}

A self-consistent set of (non-Abelian Vlasov) equations for the
distribution function and the mean color field is obtained by augmenting
the Boltzmann equation
with the Yang-Mills equations:
\be
[D_\nu F^{\nu\mu}]^a(x) = J^{\mu\, a}(x)\ .
\ee
The covariant derivative is defined as $D_\mu^{ac} =\pa_\mu \delta^{ac} +
g\, f^{abc} A_\mu^b$. The total color
current $J^{\mu\,a}(x)$ is given by the sum of all contributions
from helicities and particle species,
\be
J^{\mu\, a}(x) = \sum_{\rm helicities}\sum_{\rm species}\,\, j^{\mu\, a}(x)
= 2\ \sum_{\rm species}\,\, j^{\mu\, a}(x)\ .
\label{sumsum}
\ee
Each $j^{\mu\,a}(x)$ (spin and species indices are implicit)
is computed from the corresponding distribution function:
\begin{equation}
j^{\mu\,a} (x) = g\, \int dPdQ\ p^\mu Q^a f(x,p,Q)
\label{cr5}
\end{equation}
and it is covariantly conserved
\begin{equation}
\left(D_{\mu} j^{\mu} \right)^a (x) = 0\ ,
\label{cr6}
\end{equation}
as can be checked by using the Boltzmann equation. For later
convenience,
we define the total and individual current momentum-densities:
\be
J^{\mu\, a}(x,p) = \sum_{\rm helicities}
 \sum_{\rm species} j^{\mu\, a}(x,p)\ ,\qquad
j^{\mu\,a} (x,p) =  g\,\int dQ\ p^\mu Q^a f(x,p,Q)\ .
\label{cudens}
\ee

The above framework is now employed to study the soft
excitations in a hot, color-neutral quark-gluon plasma.
In the high-temperature limit, the masses of the particles
can be neglected and shall henceforth be taken to vanish.
The wavelenght of  the soft
excitations is of order 1/$g\,|A|$, and the coupling constant $g$
is assumed to be small.
We then expand the distribution function $f(x,p,Q)$ in powers of $g$:
\be
f=f^{(0)}+gf^{(1)}+g^2f^{(2)}+...\ ,
\label{L1}
\ee
where $f^{(0)}$ is the
equilibrium distribution function in the absence of a net color
field, and is given by:
\be
f^{(0)}(p_0)=C\ n_{B,F}(p_0)\ .
\ee
Here $C$ is a normalization constant and $n_{B,F}(p_0)=1/(e^{\beta
|p_0|}\mp 1)$ is the bosonic, resp. fermionic, probability distribution.

At leading-order in $g$, the induced current (\ref{cudens})
is
\be
j^{\mu\,a}(x,p)=g^2 \int dQ\  p^{\mu} Q^a f^{(1)}(x,p,Q)\ ,
\label{L2}
\ee
while the Boltzmann equation (\ref{boltzmann}) reduces to
\be
p^{\mu} \left({\pa\over\pa x^{\mu}}-g\, f^{abc} A_{\mu}^b Q_c
{\pa\over\pa Q^a}\right)
f^{(1)}(x,p,Q) = p^{\mu} Q_a F_{\mu \nu}^a {\pa\over \pa p_{\nu}}
f^{(0)}(p_0)\ .
\label{L3}
\ee
Due to the softness of the excitation, the $\pa_x$ term in the above equation
is of order $g\,|A|$, so we are taking into account consistently all
contributions in order $g$.

The equations (\ref{L2}) and (\ref{L3}) yield
the following constraint on the induced current:
\be
[\,p \cdot D\,\, j^{\mu}(x,p)]^a = g^2\, p^{\mu} p^{\nu} F_{\nu \rho}^b
{\pa\over\pa p_{\rho}} \left(\int dQ\ Q^a Q_b f^{(0)}(p_0)\right) \ ,
\label{L4}
\ee
where, from color symmetry, we expect that
$\int dQ\ Q^a Q_b f^{(0)}(p_0)=C_{B,F}\,\, n_{B,F}(p_0)\,
\delta^a_{\,b}$ with $C_{B}=N,\ C_F={1\over 2}$ for gluons, resp. fermions.
Thus, upon summation over all species ($N_F$ quarks, $N_F$ antiquarks and
one [$(N^2-1)$-plet] gluon)
and helicities (2 for quarks-antiquarks and massless gluon),
(\ref{L4}) yields,
\be
[\,p \cdot D\,\, J^{\mu}(x,p)]^a = 2\,g^2\, p^{\mu} p^{\nu}
F_{\nu 0}^a(x) {d \over dp_0}[N\, n_B(p_0)+N_F\, n_F(p_0)] \ .
\label{L5}
\ee
This equation is equivalent\footnote{Our notational
conventions differ from those of \cite{JLL}: the
$\delta(p^2)$-function appearing in (3.13) there is here contained in the
(massless limit of the) momentum measure $dP$ (\ref{measurep}).}
to the condition (3.13) of \cite{JLL} for the induced current
(this condition has been previously obtained in \cite{BI}).

The subsequent steps which lead to the gauge
invariance condition for HTL's can be found in \cite{JLL}.
Assuming that the induced current can be expressed as the functional
derivative of an effective action allows one to derive, from (\ref{L5}), a
condition on the latter action. The result can be presented as (eq. (3.26)
in \cite{JLL}):
\be
\pa_+\,{\delta W(A_+)\over \delta A_+}
+g\,\left[ A_+,{\delta W(A_+)\over \delta A_+}\right]
=4\sqrt{2}\,\pi^3\,m_D^2\, \pa_0 A_+\ ,
\label{3.26}
\ee
where $\ \pa_+\equiv Q^\mu_+\,\pa_\mu$,
$\ A_+\equiv Q^\mu_+\, A_\mu$, with
$\ Q^\mu_+={1\over\sqrt{2}}(1,{{\bf p}\over p_0})\ $,
and $\ m_D=gT\,\sqrt{{N+N_F/2\over 3}}\ $ is
the Debye screening mass.
The effective action generating the HTL's has the form (first derived in
\cite{BPFT}):
\be
\Gamma_{\rm HTL} ={m_D^2\over 2}\,\int d^4\!x\,A^a_0(x)A^a_0(x)
-\int {d\Omega\over (2\pi)^4} \,W(A_+)\ ,
\label{Gamma}
\ee
where $d\Omega$ denotes integration over all angular directions of the unit
vector ${{\bf p}\over p_0}$. The gauge invariance
condition (\ref{3.26}) has been solved for $W$ \cite{TW,EN},
thereby obtaining closed
expressions for the generating functional $\Gamma_{\rm HTL}$.
This concludes our derivation of hard thermal loops from
classical transport theory.

Let us now summarize and discuss our results. The classical transport
theory for a hot collisionless plasma of colored particles is a well-suited
formalism for studying the hot quark-gluon plasma. The recently proposed
HTL's  arise naturally in this formalism.
The spin-statistics theorem is the only ``quantum mechanical" input
we used in our analysis.
Its use is entirely justified by the conditions of temperature and density
which prevail in a hot quark-gluon plasma.
With respect to alternative descriptions our approach is much
simpler.
It also demonstrates explicitly that the HTL's of QCD are a classical
effect.

\vskip3mm
{\bf Acknowledgment}

We thank Professor R. Jackiw for suggesting this problem to us and for many
useful comments.

\end{document}